\documentclass[%
 reprint,
 amsmath,amssymb,
 aps,
 prd
]{revtex4-2}

\usepackage{graphicx}
\usepackage{dcolumn}
\usepackage{bm}
\usepackage[caption=false]{subfig}
\usepackage{bigints}
\usepackage{xcolor}

\begin{document}

\title{Corrections to the hadron resonance gas from lattice QCD \\ and their effect on fluctuation-ratios at finite density}

\author{Rene Bellwied}
\affiliation{Department of Physics, University of Houston, Houston, TX 77204, USA}

\author{Szabolcs Bors\'anyi}
\affiliation{Department of Physics, Wuppertal University, Gaussstr.  20, D-42119, Wuppertal, Germany}

\author{Zolt\'an Fodor}
\affiliation{Department of Physics, Wuppertal University, Gaussstr.  20, D-42119, Wuppertal, Germany}
\affiliation{Pennsylvania State University, Department of Physics, State College, PA 16801, USA}
\affiliation{Inst.  for Theoretical Physics, ELTE E\"otv\"os Lor\' and University, P\'azm\'any P. s\'et\'any 1/A, H-1117 Budapest, Hungary}
\affiliation{J\"ulich Supercomputing Centre, Forschungszentrum J\"ulich, D-52425 J\"ulich, Germany}

\author{Jana N. Guenther}
\affiliation{Aix Marseille Univ, Universit\'e de Toulon, CNRS, CPT, Marseille, France}

\author{S\'andor D. Katz}
\affiliation{Inst.  for Theoretical Physics, ELTE E\"otv\"os Lor\' and University, P\'azm\'any P. s\'et\'any 1/A, H-1117 Budapest, Hungary}

\author{Paolo Parotto}
\affiliation{Department of Physics, Wuppertal University, Gaussstr.  20, D-42119, Wuppertal, Germany}

\author{Attila P\'asztor}
\email[Corresponding author:]{\url{apasztor@bodri.elte.hu}}
\affiliation{Inst.  for Theoretical Physics, ELTE E\"otv\"os Lor\' and University, P\'azm\'any P. s\'et\'any 1/A, H-1117 Budapest, Hungary}

\author{D\'avid Peszny\'ak}
\affiliation{Inst.  for Theoretical Physics, ELTE E\"otv\"os Lor\' and University, P\'azm\'any P. s\'et\'any 1/A, H-1117 Budapest, Hungary}

\author{Claudia Ratti}
\affiliation{Department of Physics, University of Houston, Houston, TX 77204, USA} 

\author{K\'alm\'an K. Szab\'o}
\affiliation{Department of Physics, Wuppertal University, Gaussstr.  20, D-42119, Wuppertal, Germany}
\affiliation{J\"ulich Supercomputing Centre, Forschungszentrum J\"ulich, D-52425 J\"ulich, Germany}

\date{\today}

\begin{abstract}
The hadron resonance gas (HRG) model is often believed to correctly
describe the confined phase of QCD. This assumption is the basis 
of many phenomenological works on QCD thermodynamics and of the 
analysis of hadron yields in relativistic heavy ion collisions. 
We use first-principle lattice simulations to calculate corrections 
to the ideal HRG. Namely, we determine the sub-leading fugacity expansion 
coefficients of the grand canonical 
free energy, receiving contributions from processes like kaon-kaon or
baryon-baryon scattering. We achieve this goal by performing a two dimensional
scan on the imaginary baryon number chemical potential ($\mu_B$) 
- strangeness chemical potential ($\mu_S$) plane, where the fugacity
expansion coefficients become Fourier coefficients.
We carry out a continuum limit estimation of these coefficients 
by performing lattice simulations with temporal extents of 
$N_\tau=8,10,12$ using the 4stout-improved staggered action. 
We then use the truncated fugacity expansion to extrapolate ratios 
of baryon number and strangeness fluctuations and correlations 
to finite chemical potentials. 
Evaluating the fugacity expansion along the crossover line, 
we reproduce the trend seen in the experimental data on net-proton fluctuations by the STAR collaboration.  
\end{abstract}

\maketitle

\section{\label{sec:intro} Introduction} 

The study of the QCD phase diagram has been a very active area of research
for the last few decades. While much is known about
the thermodynamics of QCD at zero baryon number chemical 
potential, such as the temperature of the crossover 
transition~\cite{Aoki:2006we, Aoki:2006br, 
Borsanyi:2010bp, Bazavov:2011nk} and the equation of 
state~\cite{Borsanyi:2013bia, Bazavov:2014pvz, 
Borsanyi:2016ksw, Bazavov:2013pra}, the properties of 
the theory at finite baryon densities remain elusive. 
Effective models predict that the crossover transition
turns into a real phase transition at a critical 
endpoint~\cite{Fukushima:2013rx, Kovacs:2016juc,Critelli:2017oub}. However, confirmation of this
feature is needed from a first principles approach and/or experiment. 
The main goal of the currently ongoing experimental 
effort at the second Beam Energy Scan program at RHIC in  2019-2021 
is locating the supposed critical endpoint of QCD.

Direct first principle lattice simulations at finite chemical potential are hampered by the infamous sign 
problem~\cite{deForcrand:2010ys}. Methods to circumvent it include 
reweighting~\cite{Hasenfratz:1991ax,Fodor:2001au,Fodor:2004nz,Fodor:2001pe,Giordano:2020uvk,Giordano:2020roi},
Taylor expansion around zero chemical potential~\cite{Allton:2002zi,Allton:2005gk,Gavai:2008zr,Borsanyi:2011sw,
Borsanyi:2012cr,Bellwied:2015lba,Ding:2015fca,Bazavov:2017dus,Fodor:2018wul, 
Giordano:2019slo, Giordano:2019gev,Bazavov:2020bjn}, and extrapolation from purely imaginary chemical potential
~\cite{deForcrand:2002hgr,DElia:2002tig,DElia:2009pdy, 
Cea:2014xva,Bonati:2014kpa,Cea:2015cya,Bonati:2015bha,
Bellwied:2015rza,DElia:2016jqh,Gunther:2016vcp,Alba:2017mqu,
Vovchenko:2017xad,Bonati:2018nut,Borsanyi:2018grb,
Borsanyi:2020fev,Pasztor:2020dur}. The first of these 
methods has so far proved too expensive to apply on fine lattices.
Therefore, no continuum extrapolated results exist with this approach
so far. The latter two methods, on the other hand,
involve analytic continuation, which is an ill-posed
problem, regardless of whether the available data is a 
number of Taylor coefficients at zero chemical potential or the 
value of some observable at a number of points at imaginary 
chemical potential. In such a case, it 
is important to use physical insight to argue what the
functional form of a given observable could be as a function 
of the chemical potential.

The confined phase of QCD is often assumed to be well described by 
the ideal hadron resonance gas (HRG) model~\cite{Hagedorn:1965st, 
Dashen:1969ep, Dashen:1974jw, Dashen:1974yy, Venugopalan:1992hy}.
The HRG model is based on the assumption that a gas of 
interacting hadrons can be described as a gas of 
non-interacting hadrons and resonances. The inclusion of 
the resonances as free particles is an approximate 
way of taking into account resonant
interactions between the stable 
hadrons~\cite{Dashen:1974jw, Dashen:1974yy}.
The model describes bulk thermodynamic observables - like 
the pressure  or the energy density - 
obtained from first principle lattice calculations 
rather well at zero chemical 
potential~\cite{Karsch:2003vd, Huovinen:2009yb, 
Borsanyi:2010cj, Bazavov:2014pvz}. 
However, when looking at observables probing 
finite chemical potentials, namely 
Taylor expansion coefficients in the 
chemical potentials $\mu_B, \mu_S$ and $\mu_Q$ near zero
or in the respective fugacities 
$e^{\mu_B/T}, e^{\mu_S/T}$ and $e^{\mu_Q/T}$ near $1$, 
some discrepancies start to emerge between 
the HRG model and lattice calculations. 
Some of these discrepancies can be
traced back to the fact that some fugacity 
expansion coefficients with baryon number zero and one
are underestimated by the HRG model. 
Though this is not the only possibility, 
this type of discrepancy can be interpreted within the bounds of the HRG model 
itself, and has been used to try to infer 
the existence of as of yet unobserved 
hadrons~\cite{Majumder:2010ik, Bazavov:2014xya, 
Lo:2015cca, Alba:2017mqu}. The other possibility is that instead of
more resonances, a better treatment of resonances is needed, taking into
account finite widths and also non-resonant interactions~\cite{Dashen:1969ep, Dashen:1974jw, 
Dashen:1974yy, Weinhold:1997ig, Friman:2015zua, Lo:2017lym}. Of course, both statements can be true at the 
same time. For a precise description of the thermodynamics, we most likely need better
knowledge of the mass spectrum at higher energies, as well as a more accurate treatment
of resonances.

Other discrepancies between the ideal HRG and the lattice 
are impossible to resolve by supposing the existence of more resonances. 
These discrepancies can be traced back to the observation that 
even in the temperature range below the crossover, the HRG 
fails to describe sub-leading fugacity expansion coefficients.

In principle, the hadron resonance gas can be systematically 
improved by the S-matrix formulation due to Dashen, Bernstein and Ma
~\cite{Dashen:1969ep, Dashen:1974jw, Dashen:1974yy}, which allows for the calculation 
of the fugacity expansion coefficients, if enough information
is known about the scattering matrix of the hadrons. 
Applying the S-matrix formalism can lead to a better description of QCD
thermodynamics. Ref.~\cite{Lo:2017lym}
shows that the baryon-electric charge correlation $\chi^{BQ}_{11}$
is particularly sensitive to pion-nucleon scattering phase shifts and that the
inclusion of these phase shifts into a hadron gas analysis leads to
an improved description of the lattice data. This observable is somewhat special though, 
in that if isospin symmetry is assumed $|S|=1$ hyperons do not contribute at all
to $\chi^{BQ}_{11}$, and therefore it is only pion-nucleon scattering that dominates the 
non-resonant contributions.

For other observables more scattering data, e.g. information about 
baryon-baryon scattering would also be needed. This is especially the
case at finite baryon density.
Unfortunately, information on these scattering 
processes is only partially available. While the 
nucleon-nucleon elastic scattering phase shifts are known 
experimentally~\cite{Arndt:1986jb, Arndt:2007qn, Workman:2016ysf}, 
the inelastic part of the S-matrix is not known.
Even less is known about scattering between hadrons other 
than nucleons. Hyperon-nucleon and hyperon-hyperon interactions
have been studied in chiral effective 
theory~\cite{Polinder:2006zh, Haidenbauer:2013oca}. In the 
last few years, the analysis of momentum space correlations 
for hadron pairs measured in pp and p-Pb collisions has also 
been used to infer properties of hadron-hadron 
interactions~\cite{Acharya:2020asf,Acharya:2019ldv,Fabbietti:2020bfg}. There are also some
lattice results available for baryon-baryon 
scattering~\cite{Doi:2017cfx,Sasaki:2019qnh,Iritani:2018sra}, but not yet with a 
continuum extrapolation.
While these mentioned research directions show a clear 
effort from the community to learn about 
scattering between other hadrons, the 
preliminary nature of these results makes the use of the 
S-matrix formalism for the fugacity expansion 
impractical at the moment. 

One simple way to nevertheless go beyond the ideal hadron 
resonance gas is to use some kind of mean field model for 
the short range repulsion and the long range attraction 
between the baryons. Such models were compared to lattice 
results in Refs~\cite{Vovchenko:2016rkn, Alba:2016fku,
Huovinen:2017ogf, Vovchenko:2017xad}. These works in 
particular emphasized the importance of the hard core repulsive
interactions between hadrons when describing thermodynamics 
at finite baryon chemical potential. This type of interaction
is completely absent in the ideal HRG and leads to a 
sizable negative contribution to the fugacity expansion coefficients
with baryon number two. Such approaches, while interesting, are 
very far from the first principle approach that the 
S-matrix formulation could provide, were the 
necessary S-matrix elements known. 
In fact, the flavour dependence of excluded volume parameters 
used in the literature so far have been quite arbitrary,
often assuming the same excluded volume for all hadrons. We 
believe the present calculation of the fugacity expansion coefficients can lead to
the construction of more realistic models.

Going beyond equilibrium in the grand canonical ensemble, 
versions of the hadron resonance gas model have also been
used to interpret hadron yields in heavy ion collision 
experiments. This approach is colloquially referred to as 
thermal fits as they involve the estimation of the temperature 
and chemical potential where the yields of hadrons are 
frozen, the so-called chemical freeze-out conditions.
This approach was successful in describing hadron 
yields ~\cite{Cleymans:1998yb, Andronic:2008gu, 
Andronic:2017pug,Manninen:2008mg,  Abelev:2009bw, Aggarwal:2010pj}, which is quite remarkable, considering that these yields 
at a single collision energy span many orders of magnitude.
Though an important underlying assumption here is the equilibration of 
the system produced in heavy ion 
collisions~\cite{Rapp:2000gy, BraunMunzinger:2003zz, NoronhaHostler:2007jf}, the fact
that the fits work also provides some evidence for this assumption.
In this context, it has been realized that including the pion-proton phase shifts
in the analysis changes the predicted yields as compared to the
ideal HRG at LHC energies~\cite{Andronic:2018qqt}. Of course, for consistency, 
one should extend such and S-matrix treatment
to strange hadrons as well. In lack of the necessary scattering 
data, this extension to strangeness is not straightforward~\cite{Cleymans:2020fsc}.
The inclusion of these corrections is important to precisely test the assumption of 
a single freeze-out temperature. As a competitor to this assumption, in 
the context of ideal HRG, it was shown~\cite{Flor:2020fdw}
that different freeze-out temperatures for light and strange hadrons, can significantly 
improve the description of the experimental yields at LHC and the highest RHIC energies.

Since comparisons with the available lattice data suggest that the 
agreement between full QCD and the ideal hadron resonance 
gas model gets worse at finite chemical potential, 
we suspect that non-resonant scattering effects will be even more
important at the RHIC Beam Energy Scan and future experiments 
at lower collision energies, like FAIR and NICA. 

In this work we calculate sub-leading fugacity expansion coefficients 
with first principle lattice simulations. To this end we 
perform simulations at imaginary chemical potentials,
where the fugacity expansion coefficients turn into Fourier
coefficients in the imaginary values of the chemical potentials.
This correspondence was already exploited in our earlier works.
In Ref.~\cite{Alba:2017mqu} we made a detailed analysis of the
fugacity expansion coefficients already appearing in the Boltzmann approximation
of the ideal HRG, to infer the existence of not yet 
discovered strange hadrons. In Ref.~\cite{Vovchenko:2017xad},
some of us used the fugacity expansion to emphasize the importance of
repulsive baryonic interactions near the crossover region.
In Ref.~\cite{Bellwied:2019pxh} we compared the fugacity expansion
with the Taylor expansion in the chemical potentials for 
cross-correlators of conserved charges. Here we go beyond our
earlier works by performing lattice simulations on a 2 
dimensional grid in the purely imaginary $(\mu_B^I,\mu_S^I)$
plane. This allows us for the first time to separate 
the scattering contributions to QCD thermodynamics
by the net strangeness quantum number of the participants.
In addition to giving insight on the origin of the discrepancies between full
QCD and the ideal HRG model, we believe our results on the fugacity expansion
coefficients will also be useful to tune the parameters 
of the freeze-out models in heavy ion phenomenology.

We also use the truncated fugacity expansion to 
extrapolate experimentally measured ratios of baryon 
number and strangeness susceptibilities to finite 
baryon chemical potentials on the phenomenologically 
relevant strangeness neutral line. This provides an alternative
extrapolation procedure to the standard Taylor method.
When we extrapolate on the crossover line at strangeness 
neutrality, these sub-leading coefficients 
approximately reproduces the trend seen in the experimental
data of the STAR collaboration of net-proton fluctuations~\cite{Adam:2020unf,
Nonaka:2020crv, Pandav:2020uzx}. 

The structure of the paper is as follows.  In the next section, we introduce
the basic notation and observables used in our study.  In Sec.~\ref{sec:lat} we
discuss our lattice setup. In Sec.~\ref{sec:fit} we discuss our fitting
procedure for the sectors and we present the fugacity expansion coefficients.
In Sec.\ref{sec:pheno} we calculate the fluctuation ratios using the fugacity
expansion and extrapolate to small finite density.  Finally in
Sec.~\ref{sec:con} we give a brief summary and outlook for future work.

\section{\label{sec:gce}QCD in the grand canonical ensemble} 
\subsection{Susceptibilities and the Taylor expansion} 
There is a conserved charge corresponding to each quark flavor of QCD. Working
with three flavors, the grand canonical partition function can be then written
in terms of three quark number chemical potentials $\mu_u,\mu_d$ and $\mu_s$.
The generalized susceptibilities are defined to be 
derivatives of the grand potential (or pressure)  with respect to these chemical potentials:
\begin{equation}
\chi^{uds}_{ijk}= \frac{\partial^{i+j+k} (p/T^4)}{
(\partial \hat\mu_u)^i
(\partial \hat\mu_d)^j
(\partial \hat\mu_s)^k
} \, \, ,
\label{eq:chiquark}
\end{equation}
with the dimensionless chemical potentials $\hat\mu_X=\mu_X/T$. 
For the purpose of hadronic phenomenology it is more convenient to work with
the conserved charges  $B$ (baryon number), $Q$ (electric charge) and $S$
(strangeness) instead, with chemical potentials $\mu_B,\mu_Q$
and $\mu_S$, respectively.  The basis of $\mu_u,\mu_d,\mu_s$ can be transformed
into a basis of $\mu_B,\mu_Q,\mu_S$ with a simple linear transformation, whose
coefficients are given by the $B$, $Q$ and $S$ charges of the individual
quarks:
\begin{eqnarray}
\mu_u&=&\frac{1}{3}\mu_B+\frac{2}{3}\mu_Q\,,\\
\mu_d&=&\frac{1}{3}\mu_B-\frac{1}{3}\mu_Q\,,\\
\mu_s&=&\frac{1}{3}\mu_B-\frac{1}{3}\mu_Q-\mu_S\,.
\end{eqnarray}
Analogously to the case of the quark number chemical potentials, the susceptibilities are then defined as 
\begin{equation}
\chi^{BQS}_{ijk} = \frac{\partial^{i+j+k} \left( p/T^4 \right)}{\partial \hat{\mu}_B^i \partial \hat{\mu}_Q^j \partial \hat{\mu}_S^k} \, \, .
\label{eq:chiBQS}
\end{equation}
It is straightforward to express the susceptibilities defined in Eq.~\eqref{eq:chiBQS} in terms of the 
coefficients in Eq.~\eqref{eq:chiquark} \cite{Bernard:2004je,Bazavov:2012jq,Bellwied:2015lba}. 
The susceptibilities at $\mu_B=\mu_S=\mu_Q=0$ are (up to a trivial factorial factor) the Taylor 
expansion coefficients of the pressure near that point. Due to charge
conjugation symmetry, only the even derivatives contribute. In the
present study, we always take $\mu_Q=0$ and only consider
derivatives with respect to $\mu_B$ and $\mu_S$. The Taylor expansion
therefore reads:
\begin{equation}
    \frac{p}{T^4} = \sum_{i=0}^{\infty} \sum_{j=0}^{\infty} \frac{1}{i! j!} \chi^{BS}_{ij} \hat{\mu}_B^i \hat{\mu}_S^j \rm{,}
\end{equation}
where $\chi^{BS}_{00}$ is just the dimensionless pressure 
at zero chemical potential.

We note that the Taylor expansion is probably the most natural expansion to work within the plasma phase of QCD.
As an exhibit of this, the pressure in the Stefan-Boltzmann (or infinite temperature) limit reads:
\begin{equation}
    \frac{p}{T^4} = \frac{8 \pi^2}{45} + \frac{7\pi^2}{60} N_f  + \frac{1}{2} \sum_f \left( \frac{\mu_f^2}{T^2} + \frac{\mu_f^4}{2 \pi^2 T^4} \right)
\end{equation}
In this approximation, all derivatives above 4th order are zero, and therefore the Taylor expansion is rapidly convergent.
Calculating corrections to this free gas behavior in resummed perturbation theory leads to a non-zero, but small, 
value for the sixth-order derivatives~\cite{Haque:2014rua}, leaving the qualitative conclusion of the fast convergence of 
the Taylor series in the plasma phase unchanged.

\subsection{Fugacity expansion of the free energy} 
An alternative to the Taylor expansion discussed in the previous subsection is a Laurent expansion
in the fugacity parameters $e^{\hat{\mu}_B}$ and $e^{\hat{\mu}_S}$ near $1$. Due to charge conjugation symmetry,
a combination $e^{m \hat{\mu}_B + n \hat{\mu}_S}$ and its reciprocal have the same expansion coefficients,
making the Laurent expansion an expansion in hyperbolic cosines:
\begin{equation}
    \label{eq:fugacity_expanion}
    P(T,\hat{\mu}_B,\hat{\mu}_S) =\sum_{j,k} P^{BS}_{jk}(T) \cosh(j\hat{\mu}_B-k\hat{\mu}_S) \, \,  \rm{.}
\end{equation}
The coefficients $P^{BS}_{jk}$ are also called fugacity expansion or sector coefficients, alluding to the
fact that they get contributions from the Hilbert 
subspace corresponding to the fixed values of the conserved charges
$B=j$ and $S=k$.
In the ideal HRG model, the expansion coefficients 
$P^{BS}_{00},P^{BS}_{01},P^{BS}_{10},P^{BS}_{11},P^{BS}_{12},P^{BS}_{13}$
all get sizable contributions from known hadrons and hadron resonances. 
In the Boltzmann approximation to the 
ideal HRG, coefficients like $P^{BS}_{20}$ are zero, while in 
the full HRG they are non-zero, but very small in magnitude and 
essentially negligible. 

At purely imaginary chemical potentials $\mu_q = i \mu_q^I$, 
where the sign problem is absent and lattice simulations can be performed, we have a Fourier expansion of the form:
\begin{equation}
    P(T,\hat{\mu}_B^I,\hat{\mu}_S^I) =\sum_{j,k} P^{BS}_{jk}(T) \cos(j\hat{\mu}_B^I-k\hat{\mu}_S^I) \, \,  \rm{.}
\end{equation}
Differentiation with respect to the original chemical potentials $\mu_B=i\mu_B^I$ and $\mu_S=i\mu_S^I$ gives:
\begin{eqnarray}
    \operatorname{Im} \chi^{BS}_{10} &= \sum_{j,k} j    P^{BS}_{jk}(T) \sin(j\hat{\mu}_B^I-k\hat{\mu}_S^I) \\
    \operatorname{Im} \chi^{BS}_{01} &= \sum_{j,k} (-k) P^{BS}_{jk}(T) \sin(j\hat{\mu}_B^I-k\hat{\mu}_S^I) \\
                      \chi^{BS}_{20} &= \sum_{j,k} j^2  P^{BS}_{jk}(T)   \cos(j\hat{\mu}_B^I-k\hat{\mu}_S^I) \\
                      \chi^{BS}_{11} &= \sum_{j,k} (-jk)  P^{BS}_{jk}(T) \cos(j\hat{\mu}_B^I-k\hat{\mu}_S^I) \\
                      \chi^{BS}_{02} &= \sum_{j,k} k^2  P^{BS}_{jk}(T)   \cos(j\hat{\mu}_B^I-k\hat{\mu}_S^I) \, \,  \rm{.}
\end{eqnarray}
These formulas and the higher order derivatives of 
these  will be used in our fitting procedure, to be described 
in Section~\ref{sec:fit}.

\subsection{The hadron resonance gas and its extensions}

In the ideal HRG model the free energy (or pressure) is written as
a sum of ideal gas contributions of all known hadronic resonances $H$:
\begin{equation}
    \frac{p}{T^4} = \frac{1}{T^4} \sum_H p_H = \frac{1}{VT^3} \sum_H \ln {\cal Z}_H (T, \vec{\mu}) \, \, ,
\end{equation}
with:
\begin{equation}
    \label{eq:ideal_gas_lnZ}
    \ln {\cal Z}_H = \eta_H \frac{V d_H}{2 \pi^2 T^3} \bigintssss_0^\infty \! \! \! \! dp \, p^2 \log \left[ 1 - \eta_H z_H \exp \left( - \epsilon_H/T \right) \right] \, \, ,
\end{equation}
where the subscript $H$ indicates dependence on the specific hadron or hadron resonance in the sum. The relativistic energy is $\epsilon_H = \sqrt{p^2 + m_H^2}$, where $m_H$ is the mass of the given hadron. The fugacity is $z_H = \exp\left( \mu_H /T \right)$, where the chemical potential associated to $H$ is $\mu_H = \mu_B B_H + \mu_Q Q_H + \mu_S S_H$, and the conserved charges $B_H$, $Q_H$ and $S_H$ are the baryon number, electric charge and strangeness, respectively. $d_H$ is the spin degeneracy, and the factor $\eta_H$ is $1$ for (anti)baryons (fermions) and $-1$ for mesons (bosons).

In the HRG model, the $\chi^{BQS}_{ijk}$
susceptibilities of Eq.~(\ref{eq:chiBQS}) can be expressed as:
\begin{equation}
\chi^{BQS}_{ijk} \left(T, \hat{\mu}_B, \hat{\mu}_Q, \hat{\mu}_S \right) = \sum_H B_H^i \, Q_H^j \, S_H^k \,  I^H_{i+j+k}  \, \, ,
\end{equation}
where the phase space integral at order $i+j+k$ reads:
\begin{equation}
I^H_{l} \left(T, \hat{\mu}_B, \hat{\mu}_Q, \hat{\mu}_S \right) = \frac{\partial^{l} p_H/T^4}{\partial \hat{\mu}_H^{l}} \, \, .
\end{equation}

The fugacity expansion coefficients $P^{BS}_{00}$, $P^{BS}_{01}$, $P^{BS}_{10}$, $P^{BS}_{11}$, $P^{BS}_{12}$ and 
$P^{BS}_{13}$ can be obtained via the expansion of equation~\eqref{eq:ideal_gas_lnZ} in terms of the modified Bessel functions $K_2$:
\begin{equation}
    \label{eq:Bessel}
    \ln {\cal Z}_H = \frac{V T m_H^2 d_H}{ 2 \pi^2} \sum_{n=1}^\infty \frac{(-\eta_H)^{n+1} z_H^n}{n^2} K_2\left( \frac{n m_H}{T} \right).
\end{equation}
The Boltzmann approximation consists of taking only the $n=1$ term in the above expansion, which accounts for the lowest order in the
fugacity parameters. In the Boltzmann approximation, the sectors read:
\begin{equation}
    P^{BS}_{jk} = \sum_H \delta_{B_H,j} \delta_{S_H,k} \frac{d_H m_H^2 }{ 2 \pi^2 T^2} K_2\left( \frac{m_H}{T} \right).
\end{equation}
In the full ideal HRG, a hadron with $B_H=1$ and $S_H=0$ will also give contributions to the higher
order sectors, such as $P^{BS}_{20}$ and $P^{BS}_{30}$, due to
the terms $n=2$ and $n=3$ in Eq.~\eqref{eq:Bessel}, respectively.
These are, however, exponentially suppressed due to the
behavior of the Bessel function $K_2(x) \sim \sqrt{\frac{\pi}{2 x}} e^{-x}$ 
as $x \to \infty$. These contributions are orders of magnitude
smaller than the full weight of the respective sectors as 
obtained from the lattice.

The HRG model is an approximation of the more general formula 
by Dashen, Bernstein and Ma, which gives the fugacity expansion 
coefficients in terms of the S-matrix: 
\begin{equation}
    \begin{aligned}
    P^{BS}_{jk} = &\frac{1}{\pi^3 T^3} \int_{M^{BS}_{jk}}^\infty dE E^2 K_2\left( \frac{E}{T}\right)  \\
    &\frac{1}{4i} \operatorname{Tr}_{B=j, S=k} \left( S^\dagger \frac{dS}{dE} - \frac{dS^\dagger}{dE} S\right)_c \rm{,}
    \end{aligned}
\end{equation}
where $M^{BS}_{jk}$ is the mass threshold for the $B=j$, $S=k$ channel,
the trace is taken over this Hilbert subspace, and the subscript $c$
signifies that only connected S-matrix elements are to be taken.
For the specific case of elastic $2 \to 2$ body scattering,
\begin{equation}
    \begin{aligned}
        \frac{1}{4i} \operatorname{Tr}_{B=j, S=k} \left( S^\dagger \frac{dS}{dE} - \frac{dS^\dagger}{dE} S\right)_c \rightarrow \\
    \sum_J (2J+1) \left( \frac{d \delta^{J,I=0}}{d E} + 3 \frac{d \delta^{J,I=1}}{d E} \right)\rm{,}
    \end{aligned}
\end{equation}
where the $\delta^{J,I}$ are the scattering phase shifts for angular momentum $J$ and isospin $I$ 
and the isospin singlet and triplet contributions have been written separately.
After integration by parts with respect to $E$, we get to the conclusion
that the contribution of elastic scattering is given by the 
integral of the phase shift with an exponential weight. This leads 
to the expectation that dominantly repulsive interactions will lead
to a negative sub-leading fugacity expansion coefficient. This fact
was exploited when constructing repulsive core HRG models 
and comparing them with lattice data in Refs.~\cite{Huovinen:2017ogf, Vovchenko:2017xad}.
It is also reasonable to expect, due to the exponential suppression of 
the $K_2$ Bessel functions, that in the hadronic phase there will be
a strong hierarchy of the fugacity expansion coefficients with 
increasing quantum numbers, so e.g. $P^{BS}_{01} \gg P^{BS}_{02} \gg P^{BS}_{03}$ as well 
as $P^{BS}_{10} \gg P^{BS}_{20} \gg P^{BS}_{30}$ 
and $P^{BS}_{11} \gg P^{BS}_{21} \gg P^{BS}_{31}$ etc.
It is thus a reasonable expectation that, in the hadronic phase, the fugacity
expansion will converge faster than the Taylor expansion.  This is the opposite
situation as in the plasma phase, where the Taylor expansion converges quickly,
while the fugacity expansion converges slowly. This makes the fugacity expansion
particularly useful for modelling the hadronic phase, and therefore also for
the study of chemical freeze-out in heavy ion collisions.

Here we do not utilize any S-matrix formula or any mean field 
approximation thereof, but rather calculate the sub-leading 
sector coefficients $P^{BS}_{20}$, $P^{BS}_{21}$, $P^{BS}_{22}$,
$P^{BS}_{02}$ etc. directly from lattice simulations.

\section{\label{sec:lat}Lattice setup} 
    
\begin{figure*}[t!]
    \begin{center}
        \includegraphics[width=0.48\linewidth]{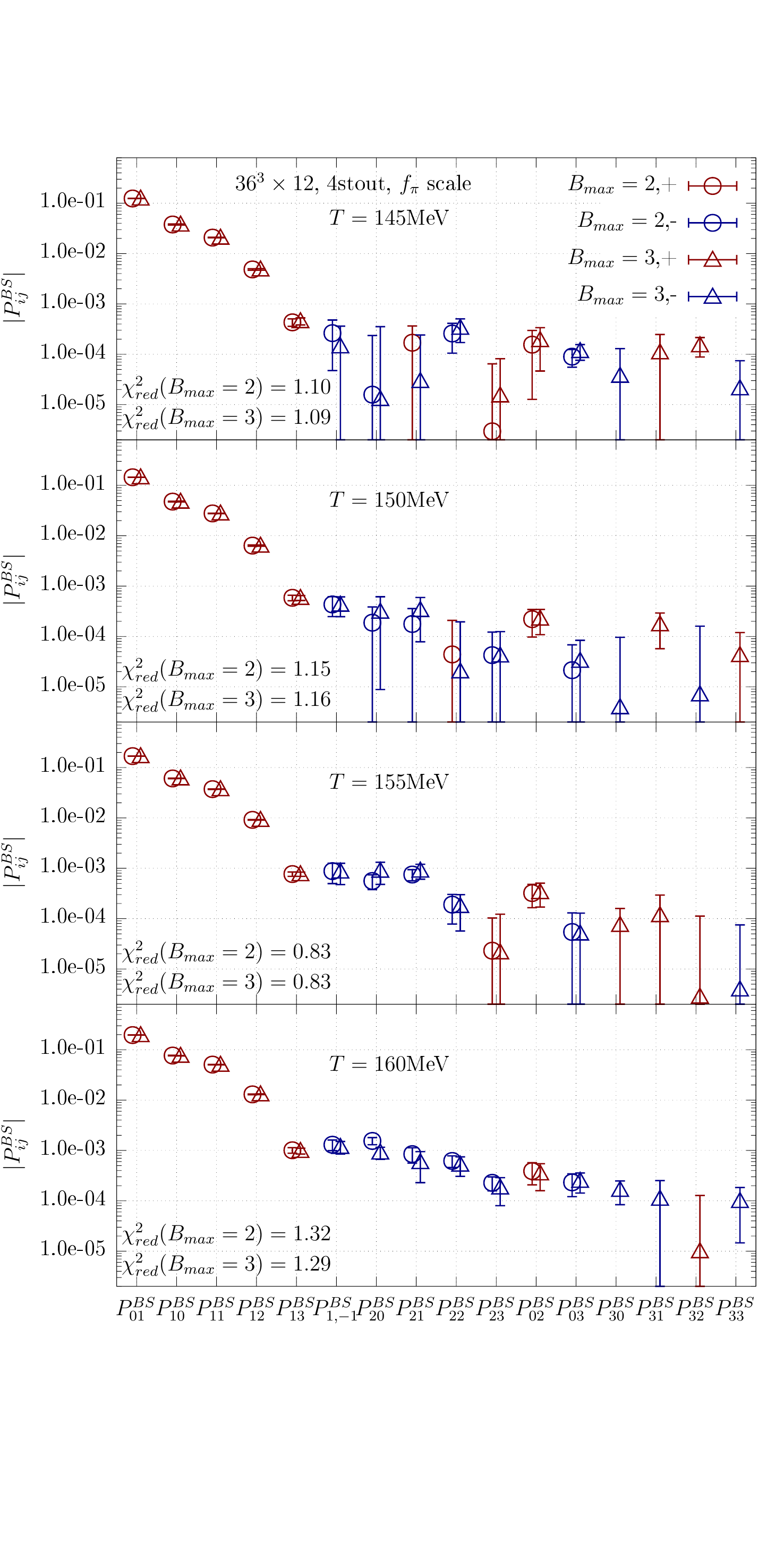}
        \includegraphics[width=0.48\linewidth]{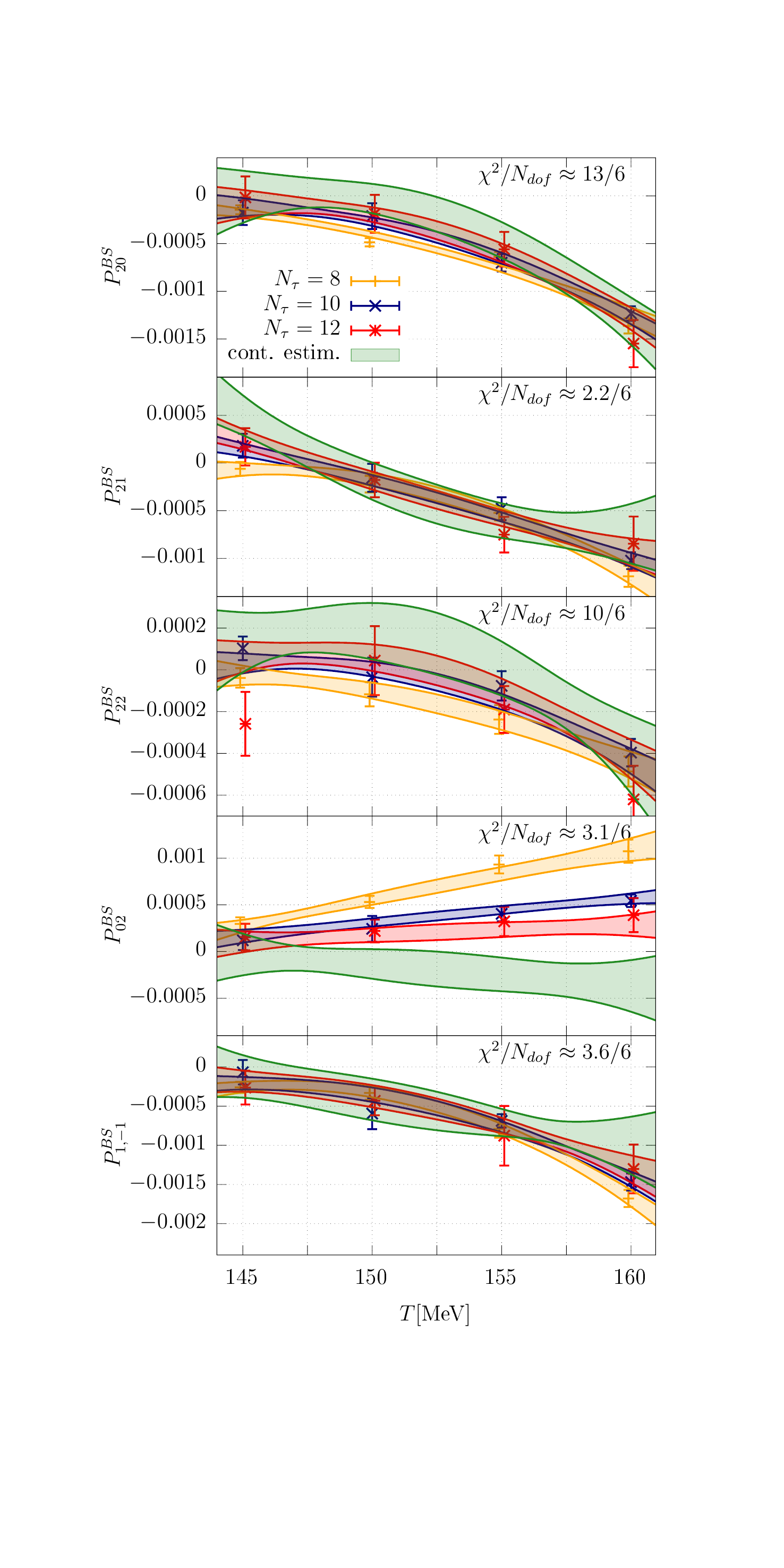}
    \end{center}
    \vspace{-3cm}
    \caption{
    \label{fig:sector_fits}    
Left: The four panels refer to four temperatures, each showing the obtained
coefficients of the fugacity expansion on a logarithmic scale (negative values
are shown in blue). Only the leftmost five are accounted for by ideal HRG. The
next seven appear in the next order, which we use later for phenomenology. For
the next order (last four coefficients with $B=3$) we see no stable signal.
There are two symbols per coefficient, triangles for the complete
fit and circles for the case without the $B=3$ part. The data refer to our finest lattice spacing.
Right: Example for the combined continuum extrapolation of the extracted
coefficients.
}
\end{figure*}

We use a staggered fermion action with 4 steps of stout
smearing~\cite{Morningstar:2003gk} with the smearing parameter $\rho=0.125$ and
a tree-level Symanzik-improved gauge action. This combination 
was first used in Ref.~\cite{Bellwied:2015lba}, where information about the
line of constant physics can be found. For the scale setting we use the pion
decay constant $f_\pi= 130.41$MeV~\cite{Tanabashi:2018oca}. 
We use lattices of temporal extent $N_\tau=8, 10$ and $12$ to perform an
estimation of the continuum value of our observables. 
The spatial extent of the lattice is given by the aspect ratio $LT\approx 3$.
Due to technical reasons, some lattices had slightly different values for this 
ratio, as given in Table.~\ref{tab:statistics}. Given the error bars on
the final results, we did not optimize this further.

For the continuum extrapolations, we assume a linear scaling in $1/N_{\tau}^2$.
Since taste breaking effects are still rather large on these lattices, we only
call our results continuum estimates, as opposed to fully controlled continuum
extrapolations, e.g. when $N_\tau=16$ is part of the extrapolation.

\begin{table}[t!]
\begin{center}
\begin{tabular}{|cc|c|c|c|c|}
\hline
    &&145~MeV&150~MeV&155~MeV&160~MeV\\
\hline
$40^3\times12$& $\mu^I=0$&
10348&10520&10345&11611\\
$32^3\times10$& $\mu^I=0$&
8518&8461&1695&9174\\
$24^3\times8$& $\mu^I=0$&
40247&39996&19953&20015\\
\hline
$36^3\times12$& $\mu^I\ne0$&
146968&154479&153513&144169\\
$32^3\times10$& $\mu^I\ne0$&
124915&81814&300779&264647\\
$24^3\times8$& $\mu^I\ne0$&
184896&171224&166034&161454\\
\hline
\end{tabular}
\end{center}
\caption{\label{tab:statistics}
Number of evaluated configurations on the various lattices and temperatures.
The $\mu^I\ne0$ statistics is distributed over 143 pairs of imaginary
strange and baryon chemical potentials.
}
\end{table}

For all values of $N_\tau$, we use simulations at four different temperatures
$T=145~\rm{MeV},150~\rm{MeV},155~\rm{MeV}$ and $160~\rm{MeV}$.  At each temperature
and each lattice spacing, we perform a two-dimensional scan in the imaginary
chemical potentials $\mu_B^I$ and $\mu_S^I$, with the chemical potentials
taking the values $(\mu_B^I,\mu_S^I) = \frac{\pi}{8} (i,j)$, with
$i=0,1,\dots,15$ and $j=0,1,\dots,9$, for a total of $9 \times 16 = 144$ simulation
points. In each $\mu_i\ne0$ point, we simulated one Rational Hybrid Monte Carlo
stream with several thousand trajectories, evaluating every fifth configuration
for the fluctuation observables as detailed in Ref.~\cite{Bellwied:2015lba}. 
Our statistics is summarized in Table~\ref{tab:statistics}. 

The statistical errors are calculated using the jackknife method. 
The estimation of the systematic errors is a more elaborate process.
Ambiguities appear at various points of the analysis, e.g. in the way
the continuum extrapolation is calculated, or how many fit parameters
we use for the extraction of the fugacity expansion coefficients.
We consider all combinations of the possibilities and take the spread
of the results as systematic error.

\section{\label{sec:fit}Fugacity expansion coefficients} 

The estimation of the coefficients $P^{BS}_{ij}$ proceeds through a 
correlated fit. On the $\mu_B=\mu_S=0$ ensembles, the fluctuations 
$\chi^{BS}_{20}$,$\chi^{BS}_{11}$,$\chi^{BS}_{02}$,$\chi^{BS}_{40}$,
$\chi^{BS}_{31}$,$\chi^{BS}_{22}$,$\chi^{BS}_{13}$ and $\chi^{BS}_{04}$ 
are included,
while for the other ensembles we use 
$\operatorname{Im} \chi^{BS}_{10}$ and 
$\operatorname{Im} \chi^{BS}_{01}$.
This leads to a block-diagonal covariance matrix with one 
$8 \times 8$ block corresponding to the $\mu=0$ ensemble, and 
$143$ blocks of size $2 \times 2$, corresponding to the ensembles 
with a non-zero value of at least one of the chemical potentials.
The covariance matrix blocks are estimated by the jackknife method 
with 24 jackknife samples.
The truncation of the fugacity expansion is somewhat ambiguous, as 
there is no single small parameter in which we actually perform this
expansion. To estimate systematic errors coming from the choice of
the ansatz, we therefore perform two fits for each ensemble, for which we 
introduce the shorthand notations $B_{\rm{max}}=2$ and 
$B_{\rm{max}}=3$. 
The sectors included in the $B_{\rm{max}}=2$ analysis are:
\begin{equation}
    \begin{aligned}
        &P^{BS}_{01}\rm{,}\ &P^{BS}_{10}\rm{,} \   &P^{BS}_{11}\rm{,}\ &P^{BS}_{12}\rm{,} \\
        &P^{BS}_{13}\rm{,}\ &P^{BS}_{1,-1}\rm{,} \ &P^{BS}_{20}\rm{,}\ &P^{BS}_{21}\rm{,} \\ 
        &P^{BS}_{22}\rm{,}\ &P^{BS}_{23}\rm{,} \   &P^{BS}_{02}\rm{,}\ &P^{BS}_{03} \rm{.}
    \end{aligned}
\end{equation}
The first five of these correspond to sectors that are already present in the
ideal HRG in the Boltzmann approximation. They also set contributions from
interactions though, e.g. non-resonant pion-nucleon interactions contribute to
$P^{BS}_{10}$, while $K$-$\Lambda$ interactions contribute to $P^{BS}_{12}$.
The $P^{BS}_{1,-1}$ is an exotic (pentaquark) sector. It gets contributions for
example for valence quark content $uudd\bar{s}$ which corresponds e.g. to
$p+K^0$ scattering. The sectors $P^{BS}_{2i}$ get contributions from
baryon-baryon scattering: $P^{BS}_{20}$ from $N-N$,
$P^{BS}_{21}$ from $N-\Lambda$,
$P^{BS}_{22}$ from $N-\Xi$ or $\Lambda-\Lambda$ 
and finally $P^{BS}_{23}$ from $N-\Omega$ or $\Lambda-\Xi$. In each
case, $\Sigma$ can replace $\Lambda$.
The coefficients $P^{BS}_{02}$ and $P^{BS}_{03}$ get contributions from two- and
three-kaon scattering, respectively. The inclusion of the $P^{BS}_{03}$ sector
with the omission of  the $P^{BS}_{30}$ sector is motivated by the lower mass
threshold of 3 kaon scattering as compared to 3 baryon scattering. In addition
to these, we also performed an analysis were four sectors
with $B=3$ were added:
\begin{equation}
P^{BS}_{30}\rm{,}\ P^{BS}_{31}\rm{,} \ P^{BS}_{32}\rm{,}\ P^{BS}_{33}\,.
\label{eq:B3}
\end{equation} 
These get contributions from three-baryon scattering, with
various strangeness contents.  Our data is not yet sufficiently
accurate to obtain a reliable estimate of the 
sectors with $B=3$, and the inclusion of these sectors does not improve the 
$\chi^2$ of the fits. Whether we include these, or not, 
the results for the $B=2$ sectors remain consistent, as we
show in the left panel of Fig.~\ref{fig:sector_fits}, where the 
sector coefficients from the two different fits on the
$N_\tau=12$ lattices are shown. This way we 
demonstrate the stability of the sectors included in the $B_{\rm{max}}=2$ set.
Only at the highest temperature $T=160$MeV, and only for one sector, $P^{BS}_{20}$,
is the systematic error coming from including the $B=3$ sectors comparable to the statistical error
of the fits.

\begin{figure}[t!]
    \centering
    \vspace{-20mm}
    \hspace{-10mm}
    \includegraphics[width=0.53\textwidth]{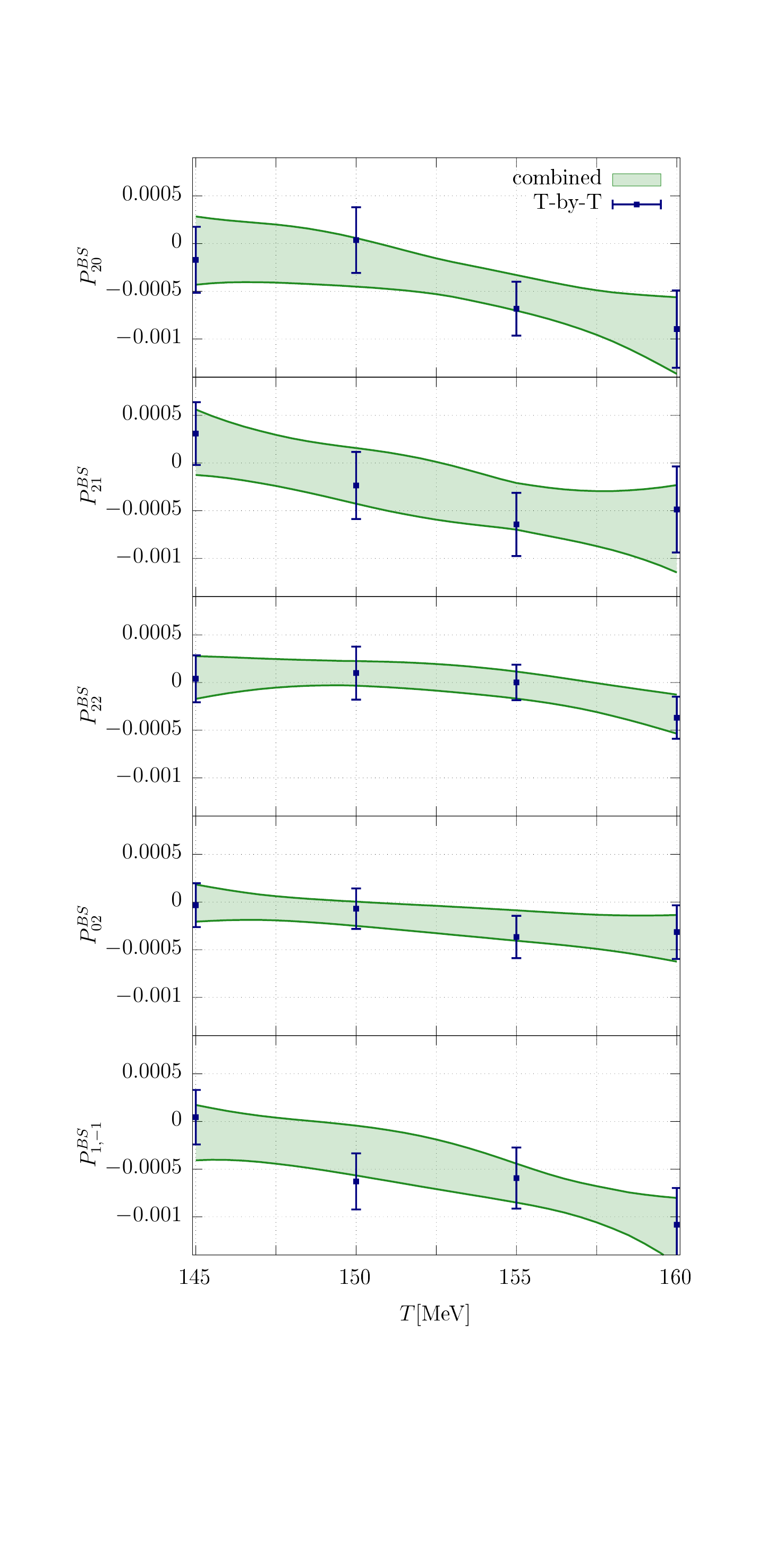}
    \vspace{-30mm}
    \caption{
        \label{fig:sector_results}
    Our continuum estimates of the beyond-ideal-HRG sector coefficients. We show the results both from our combined temperature and 
    continuum fit (green bands) and of a $T$-by-$T$ continuum limit extrapolation (blue points). Systematic errors are included, in the first case by varying $B_{\rm{max}}=2$
    vs $3$ and $b_2=0$ or $b_2\neq0$, and in the second case by varying $B_{\rm{max}}=2$ vs $3$.
    }
\end{figure}

The continuum limit estimation of the sectors proceeds through a 
combined fit in temperature and lattice spacing (or equivalently 
$N_\tau$) via the ansatz
\begin{equation}
\label{eq:continuum}
    f(T,N_\tau) = \left( a_0+a_1 T + a_2 T^2 \right) 
    + \left( b_0 + b_1 T + b_2 T^2 \right) \frac{1}{N_\tau^2}\,.
\end{equation}
For the systematic error we compare this ansatz with and without
the coefficient $b_2$.
The continuum extrapolation of the beyond-ideal-HRG 
sectors for the case of the $B_{\rm{max}}=2$ fits at
fixed $T$ and $N_\tau$ and $b_2$ kept as a free parameter, is shown 
in Fig~\ref{fig:sector_fits} (right). The other
3 fits look quantitatively similar. All of the continuum fits have 
acceptable fit quality, with $Q$ values over $1\%$. As a 
conservative 
estimate of the systematics, we combine them with uniform weights. 
As can be seen in the right panel of Fig~\ref{fig:sector_fits}, 
the slopes of the continuum extrapolations of all beyond-ideal-HRG
sectors appear to be mild, except for the sector $P^{BS}_{02}$, 
which corresponds to kaon-kaon scattering, and changes its sign 
during the continuum extrapolation. As expected, this sector - being 
related to kaons - suffers from relatively large taste-breaking 
effects.

The final results for the beyond-ideal-HRG sectors can be seen in
Fig.~\ref{fig:sector_results}. Within the statistical precision of our results,
$P^{BS}_{20}$ is roughly the same as $P^{BS}_{21}$, while $P^{BS}_{22}$ is
smaller than the previous two. As a comparison, the
ideal HRG model prediction for the sum $\sum_k P^{BS}_{2k}$ at $T=155$MeV
is of the order $10^{-5}$, orders of magnitude lower than what we 
see here. The two-kaon scattering
sector $P^{BS}_{02}$ goes slightly below zero at 
around $155$MeV within $1\sigma$ uncertainty. 
The three-kaon sector $P^{BS}_{03}$ is consistent with zero in the
entire temperature range and is therefore not included in the plot. 
An upper limit on its magnitude with $1\sigma$ uncertainty 
is $2 \cdot 10^{-3}$. The exotic sector $P^{BS}_{1,-1}$ is 
rather large, consistently with our earlier, statistically independent finding
in Ref.~\cite{Bellwied:2019pxh} on $N_\tau=12$ lattices. 
We have already published the leading sector coefficients for which 
the ideal HRG has a prediction in the Boltzmann-approximation, 
namely $P^{BS}_{01}$, $P^{BS}_{10}$, $P^{BS}_{11}$, 
$P^{BS}_{12}$, $P^{BS}_{13}$ in Ref.~\cite{Alba:2017mqu}.
We will not repeat the results for those sectors here.

\section{\label{sec:pheno}Fluctuation-ratios at finite baryon density} 

Having the coefficients of the fugacity expansion, the thermodynamics
can be readily obtained. In this section we calculate the baryon number
and strangeness fluctuations and their ratios in the studied temperature range.
There is no difficulty in evaluating Eq.~\ref{eq:fugacity_expanion}
and its $\mu_B$- and $\mu_S$-derivatives at any chemical potential. 

Heavy ion collisions involving lead or gold atoms correspond to 
the conditions $\chi^S_1 = 0$ and $\chi^B_1 = 0.4 \chi^Q_1$. 
For the purposes of the present study, we impose strangeness neutrality
and leave the second conditions for future work. In fact, we use
the simplified form $\chi^B_1 = 0.5 \chi^Q_1$, which is realized at 
vanishing electric charge chemical potential.

\begin{figure}[t!]
    \centering
    \vspace{-16mm}
    \includegraphics[width=0.48\textwidth]{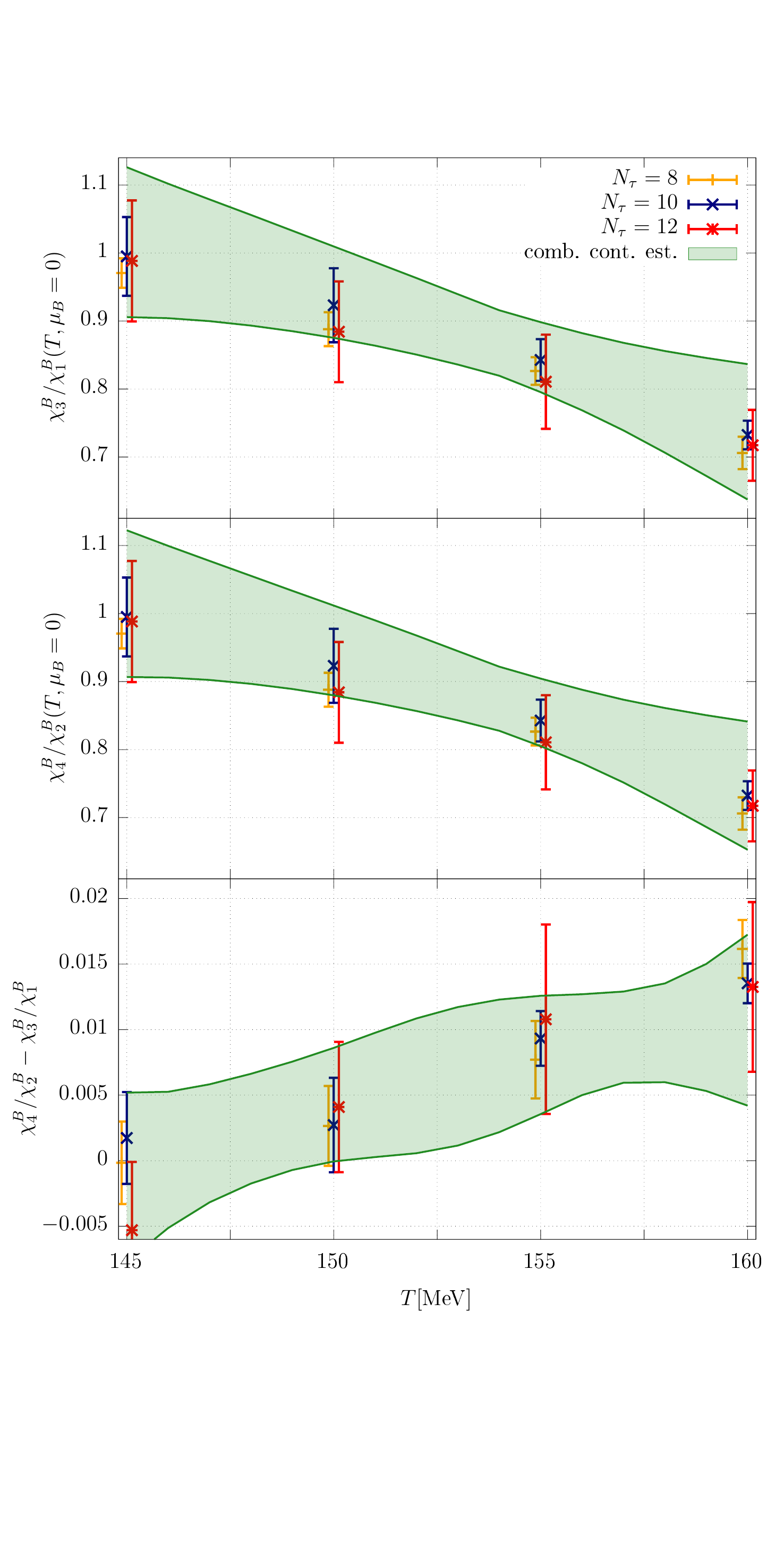}
    \vspace{-3.4cm}
    \caption{
    \label{fig:R_31_42_mu0} 
    Fluctuation-ratios $\chi^B_3 / \chi^B_1$ and $\chi^B_4 / \chi^B_2$ obtained from 
    the fugacity expansion truncated at the $B_{\rm{max}}=2$ level at $\mu_B=0$ on
    our $N_\tau=8,10$ and $12$ lattices and our continuum estimates from these data. The 
    points at finite lattice spacing include a systematic error coming from whether we 
    used the 12- or the 16-parameter fit to determine the $B_{\rm{max}}=2$ sectors. The
    continuum results include systematic error from 4 fits, in addition for the 12 vs 16
    parameter fits at fixed $N_\tau$ and $T$ we also include a 5- vs 6-parameter combined
    $T$ and $N_\tau$ continuum fit.
    }
\end{figure}

To show the magnitude of the cut-off effects, we start with a pair of quantities at $\mu_B=0$.
The fugacity expansion, and more generally imaginary chemical potential simulations offer an 
efficient way to calculate susceptibilities at $\mu_B=0$. In Ref.~\cite{Borsanyi:2018grb} we calculated
the ratios $\chi^B_3/\chi^B_1$ and $\chi^B_4/\chi^B_2$ at a finite lattice spacing of $N_\tau=12$ with the
same lattice action used here.
Here we show continuum estimates of the fluctuation ratios 
$\chi^B_3/\chi^B_1$ and $\chi^B_4/\chi^B_2$ at $\mu_B=0$, together with the data at finite $N_\tau$ 
in Fig.~\ref{fig:R_31_42_mu0}.
In the ideal HRG model $\chi^B_4/\chi^B_2=1$ for all temperatures, meaning that above $T=150$MeV
our results show a clear deviation from the HRG prediction, due to presence of the non-zero beyond-ideal-HRG 
sectors.
The difference between the two ratios $\chi^B_4/\chi^B_2-\chi^B_3/\chi^B_1$ is
also shown. In the $\mu_S=0$ case the two ratios at $\mu_B=0$ 
are identical. The difference between the two ratios comes from 
imposing the strangeness 
neutrality condition $\chi^S_1=0$.
This difference also shows mild cut-off effects. 

After the sectors are obtained, we perform extrapolations 
to real chemical potentials using the ansatz of
Eq.~\eqref{eq:fugacity_expanion} truncated at the 
$B_{\rm{max}}=2$ level. We extrapolate first at fixed $T$ and $N_\tau$.
We consider the 
fluctuation ratios $\chi^B_1/\chi^B_2$, $\chi^B_3/\chi^B_1$, $\chi^B_4/\chi^B_2$ and $\chi^{BS}_{11}/\chi^S_2$ on 
the strangeness-neutral line $\chi^S_1=0$, which determines 
$\mu_S$ as a function of $\mu_B$. While the extrapolation 
always uses the 12 sectors of the $B_{\rm{max}}=2$ level, 
the values of these sectors are taken both from the 
$B_{\rm{max}}=2$ and
$B_{\rm{max}}=3$ fits, to estimate the systematic errors. We then 
perform a continuum estimation at fixed values of $\mu_B/T$ with 
the same combined $T$ and $N_\tau$
fit as in the case of the baryon and strangeness sectors. 

\begin{figure}[t!]
    \centering
    \includegraphics[width=0.5\textwidth]{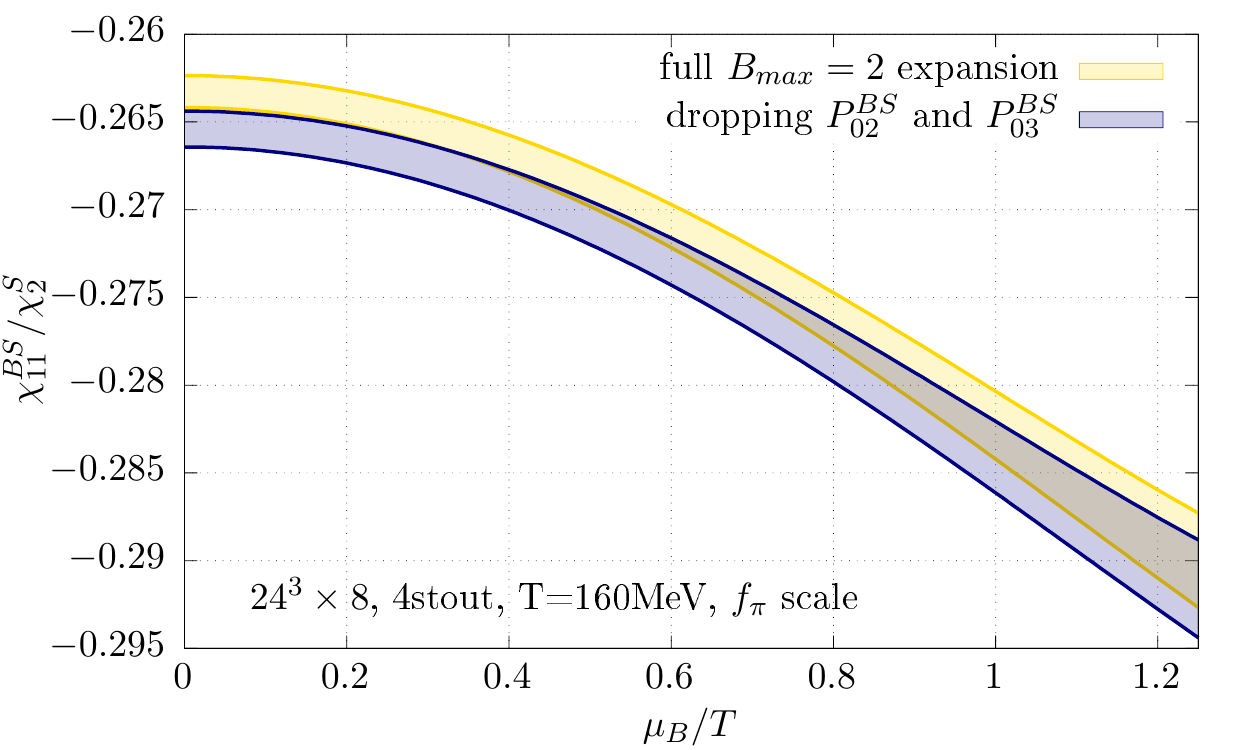}
    \caption{
    \label{fig:multi_kaon_Nt8}   
    Multi-kaon interactions have a negligible impact on the fluctuations ratios studied in this work.
    Here we show the
    effect of dropping the multi-kaon sectors from the fugacity expansion,
demonstrated on the $N_\tau=8$ data, at $T=160$MeV, where they are the largest
in our temperature range. Note also that, on the $N_\tau=8$ lattices, the
magnitude of the multi-kaon sectors is larger than our continuum estimate for
them.
    }
\end{figure}

We had one sector, $P^{BS}_{02}$, with a steep continuum extrapolation.
Should we expect additional systematic errors coming from the non-trivial
continuum scaling in the phenomenology? The answer is no, as we demonstrate in Fig.~\ref{fig:multi_kaon_Nt8}. The multi-kaon sectors do not
contribute to the baryon fluctuations. The only ratio with phenomenological
relevance where the $P^{BS}_{02}$ may be important is 
$\chi^{BS}_{11}/\chi^S_2$. We calculated this ratio with and without
the multi-kaon sectors and compared the results in Fig.~\ref{fig:multi_kaon_Nt8}
at $T=160$ MeV. Although at this temperature and this lattice spacing
we have the largest value for this difficult sector, we see hardly any
significant effect from it, especially not when compared to the statistical
errors after the continuum extrapolation step.

\begin{figure}[t!]
    \vspace{-15mm}
    \centering
    \includegraphics[width=0.5\textwidth]{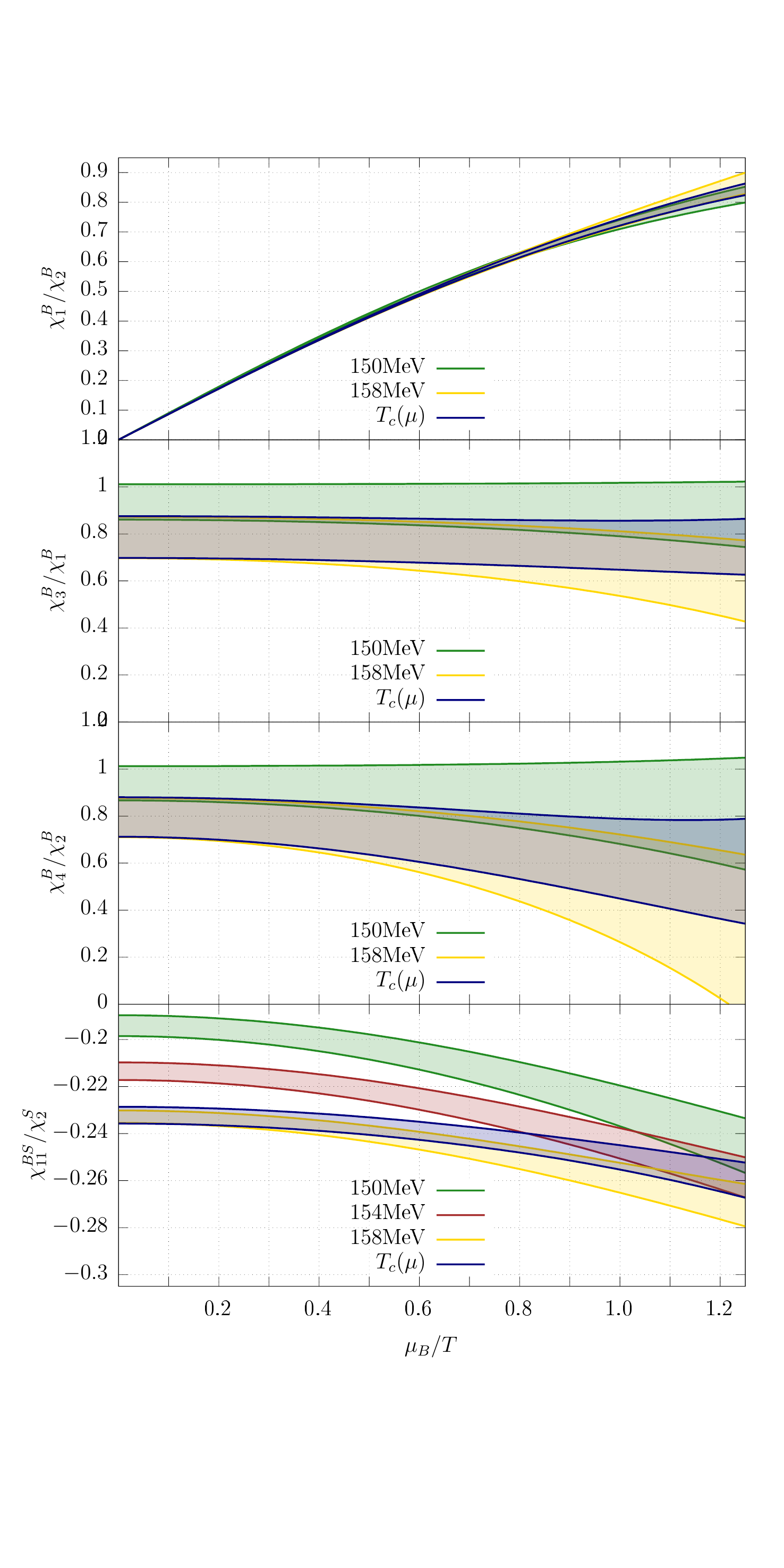}
    \vspace{-3cm}
    \caption{
        \label{fig:fluct_results}
        Continuum estimates of the fluctuation ratios 
        $\chi^B_1/\chi^B_2$, $\chi^B_3/\chi^B_1$, 
        $\chi^B_4/\chi^B_2$, $\chi^{BS}_{11}/\chi^S_2$
        from a fugacity expansion truncated at the $B_{\rm{max}}=2$ 
        level, shown as a function of the dimensionless chemical 
        potential $\mu_B/T$ for fixed temperatures, as well as
        on the crossover line $T_c(\mu_B)$. 
    }
\end{figure}

The final results for the fluctuation ratios 
$\chi^B_1/\chi^B_2$, $\chi^B_3/\chi^B_1$, $\chi^B_4/\chi^B_2$, 
$\chi^{BS}_{11}/\chi^S_2$ on the strangeness neutral line 
can be seen in Fig.~\ref{fig:fluct_results}. The first of these 
ratio is strongly dependent on the chemical potential, 
but not on the temperature,
making it a proxy of the chemical 
potential, at least for small values of $\mu_B$. 
On has to remember though that if 
the critical endpoint exists, the fluctuation $\chi^B_2$ 
diverges there, leading to $\chi^B_1 / \chi^B_2 \to 0$ at 
the critical point, and therefore making this quantity a non-monotonic function of $\mu_B$.

The other three are more strongly dependent 
on the temperature and less strongly on the 
chemical potential, therefore making them possible proxies for 
the temperature. The ratios $\chi^B_3/\chi^B_1$ and 
$\chi^B_4/\chi^B_2$
can be regarded as a baryon thermometer, while the ratio 
$\chi^{BS}_{11}/\chi^S_2$
as a strangeness-related one. This latter ratio is of large 
phenomenological interest, as experimental 
net-lambda and net-kaon fluctuations can be used to construct 
the ratio $\sigma^2_\Lambda / (\sigma^2_\Lambda+\sigma^2_K)$.
It was shown in Ref.~\cite{Bellwied:2019pxh} that this is a   
good experimental proxy of $\chi^{BS}_{11}/\chi^S_2$, not 
strongly affected by experimental effects,
which makes it a prime target for comparison with 
experiments.

We show the fluctuation ratios in Fig.~\ref{fig:fluct_results} 
as functions of the dimensionless 
chemical potential $\mu_B/T$ at a few values of the temperature,
as well as on the crossover line calculated to 
order $\mu_B^2$ in Ref.~\cite{Borsanyi:2020fev}:
\begin{equation}
    T_c(\mu_B) \approx T_c^0 \left(1- \kappa_2 \hat{\mu}_B^2 \right) \rm{,}
\end{equation}
with $T_c^0 = 158.0 \pm 0.6$ and $\kappa_2 = 0.0153 \pm 0.0018 $. 
The errors on these numbers are included in the error estimation, 
but are negligible. Note that, since the crossover temperature changes 
very little in the chemical potential range of our study, the 
$1\sigma$ bands on the $T_c(\mu_B)$ line always overlap with 
the $1\sigma$ bands for a fixed $T=T_c^0=158$MeV.

\begin{figure}[t]
    \centering
    \includegraphics[width=0.5\textwidth]{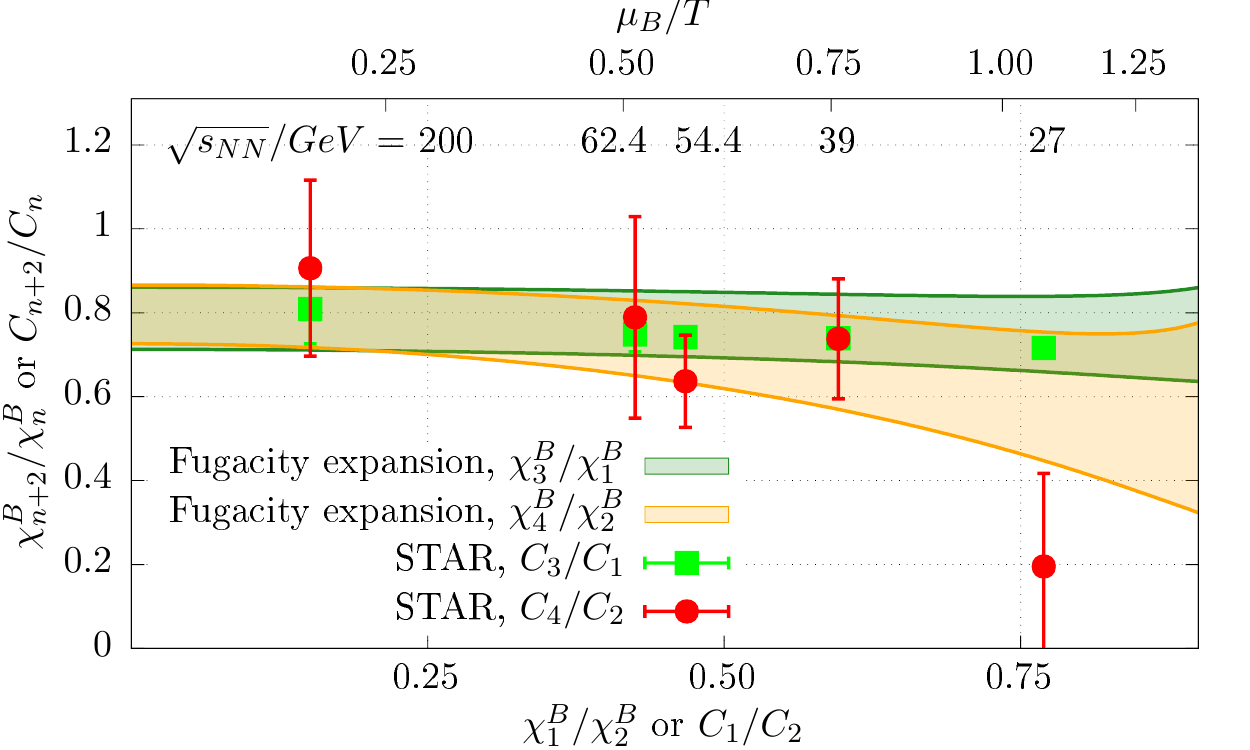}
    \caption{
\label{fig:STAR_compare}
Our continuum estimates of the fluctuation ratios $\chi^B_3/\chi^B_1$ and
$\chi^B_4/\chi^B_2$ compared with STAR data on net-proton fluctuations from
Ref.~\cite{Adam:2020unf}. Both the values of the fluctuations and the trend
as a function of baryon density are consistent.
    }
\end{figure}

Our results on the ratios $\chi^B_3/\chi^B_1$ and $\chi^B_4/\chi^B_2$ are also
shown as a function of $\chi^B_1/\chi^B_2$ in Fig.~\ref{fig:STAR_compare}.
Within $1\sigma$, our results are consistent with recent lattice results on the
susceptibility ratios using the Taylor method~\cite{Bazavov:2020bjn}.  In
Fig.~\ref{fig:STAR_compare} we also compare to the data on net-proton
fluctuations from the STAR collaboration~\cite{Adam:2020unf, Nonaka:2020crv,
Pandav:2020uzx}, the net-proton skewness-to-mean ratio $C_3 / C_1$ and the
net-proton kurtosis-to-variance ratio $C_4/C_2$, as functions of the
net-proton mean-to-variance ratio $C_1/C_2$ at chemical freeze-out. The
advantage of using these variables in the comparison is that is does not
involve any modelling of the freeze-out conditions, other than assuming that
chemical freeze-out happens close to the QCD crossover on the 
phase diagram. Our results are consistent with the experimental results. 
While such a direct comparison
suffers from many caveats~\cite{Kitazawa:2011wh, Skokov:2012ds, 
Bzdak:2019pkr, Bellwied:2019pxh, 
Braun-Munzinger:2020jbk,Vovchenko:2020tsr}, the similarity in the trends
supports the idea that experimentally observed net-proton
fluctuation ratios reflect with some accuracy the thermal
fluctuations in an equilibrated QCD medium.

\section{\label{sec:con}Summary and outlook} 

We have calculated fugacity expansion coefficients of the QCD 
pressure beyond the ideal HRG model, separating contributions 
to the QCD free energy coming from Hilbert subspaces with different
values of the baryon number and strangeness quantum numbers.
This allows one to quantify the importance of processes like 
kaon-kaon and baryon-baryon scattering, when modelling the QCD medium
in the hadronic phase, but close to the crossover.
We estimated the continuum value of these coefficients 
with lattice simulations of temporal extent
$N_\tau=8,10$ and $12$ using the staggered discretization. 
We observed large cut-off effects in the kaon and multi-kaon
sectors, only. To study these, and to make the continuum
extrapolation more robust, the inclusion of finer lattices is
desirable.

Note that our study was limited to an aspect ratio 
of $LT \approx 3$, future studies should also investigate finite
volume effects in the baryon-strangeness sectors. However,
a strong volume dependence is more likely to be observed
in correspondence with the electric charge sectors. 
The full picture of baryon interactions in the hadronic phase
will emerge from the three-dimensional mapping of the $\mu_B-\mu_S-\mu_Q$
space. In this work we restricted the space to $\mu_Q=0$.
The sectors we obtained are sums of various charge sectors,
$P^{BS}_{ij} = \sum_k P^{BSQ}_{ijk}$, and we cannot differentiate 
between the terms, though it would be possible in a more elaborate setup.
Still, the level of separation achieved in this work already
provides plenty of new information for hadronic modelling of the QCD
medium.

We used the truncated fugacity expansion to 
calculate phenomenologically relevant fluctuation ratios on the 
strangeness neutral line both as a function of the chemical potential
and as a function of the baryon number mean-to-variance ratio,
which can be regarded as a proxy of the baryo-chemical potential.
While a direct comparison is by no means trivial, the 
fugacity expansion coefficients appear to describe the trend in 
the STAR data on net-proton fluctuations~\cite{Adam:2020unf,
Nonaka:2020crv, Pandav:2020uzx}.

It has been pointed out in the literature, that the modifications 
of the ideal HRG model that include the effects of global baryon number 
conservation lead to a reduction in higher order fluctuations 
and therefore describe the experimental data better. For a 
recent study of these effects 
see~\cite{Braun-Munzinger:2020jbk}. In a recent paper, it was also 
pointed out that the decrease in the kurtosis with increasing chemical 
potential observed in the data is likely not due to critical point effects \cite{Mroczek:2020rpm}.
The corrections to ideal HRG studied in this work have a similar
magnitude as the experimental effects like canonical effects and
volume fluctuations. 
Since both baryon interactions and the
global conservation laws appear to push the $\chi^B_4/\chi^B_2$ and 
$\chi^B_3/\chi^B_1$ ratios down, it is important to have an estimate of
the relative size of these types of effects under realistic conditions.
Performing a study of this kind is an important task for the 
near future, as it will guide the correct interpretation of STAR data.

\begin{acknowledgments} 
This project was partly funded by the DFG grant SFB/TR55 and also supported by
the Hungarian National Research, Development and Innovation Office, NKFIH
grants KKP126769. 
The project leading to
this publication has received funding fromExcellence Initiative of
Aix-Marseille University - A*MIDEX, a French “Investissements d’Avenir”
programme, AMX-18-ACE-005.
    The project also received support from the BMBF
grant 05P18PXFCA. Parts of this work were supported by the National Science
Foundation under grant no. PHY1654219 and by the U.S. Department of Energy,
Office of Science, Office of Nuclear Physics, within the framework of the Beam
Energy Scan Theory (BEST) Topical Collaboration.
This research used resources of the Oak Ridge Leadership Computing Facility, which is a DOE Office of Science User Facility supported under Contract DE-AC05-00OR22725. A.P. is supported by the
J\'anos Bolyai Research Scholarship of the Hungarian Academy of Sciences and by
the UNKP-20-5 New National Excellence Program of the Ministry of Innovation and
Technology.  
RB acknowledges support from  the U.S. Department of Energy Grant No. DE-FG02-07ER41521. 
The authors gratefully acknowledge the Gauss Centre for
Supercomputing e.V. (www.gauss-centre.eu) for funding this project by providing
computing time on the GCS Supercomputer JUWELS and JURECA/Booster at J\"ulich
Supercomputing Centre (JSC). Parts of the computations were performed
on the QPACE3 system, funded by the DFG.
C.R. also acknowledges the support from the Center
of Advanced Computing and Data Systems at the University of Houston. 

\end{acknowledgments} 

\providecommand{\noopsort}[1]{}\providecommand{\singleletter}[1]{#1}%

\end{document}